\documentclass[aps,pre,amsmath,amssymb,longbibliography,superscriptaddress]{revtex4-1}
\usepackage[dvipsnames,rgb,dvips]{xcolor}
\usepackage[dvipsnames,rgb,dvips]{xcolor}
\usepackage{overpic}
\usepackage{graphicx}
\usepackage{amssymb}
\usepackage{dcolumn}
\usepackage{epstopdf}
\usepackage{siunitx}
\usepackage{mathrsfs}
\makeatletter
\def\amsbb{\use@mathgroup \M@U \symAMSb}
\makeatother
\usepackage{bbm}
\expandafter\let\csname equation*\endcsname\relax
\expandafter\let\csname endequation*\endcsname\relax 
\usepackage{diagbox}
\usepackage{tabularx}
\usepackage{amsmath}
\usepackage{amsfonts}
\makeatletter
\def\amsbb{\use@mathgroup \M@U \symAMSb}
\makeatother
\usepackage[bbgreekl]{mathbbol}
\newcommand{\ve}[1]{\ensuremath{\mbox{\boldmath$#1$}}}
\newcommand{\ma}[1]{\ensuremath{\mathbb{#1}}}

\begin{document}
	
	\title[Unsteady and inertial dynamics of an active particle in a fluid]{Unsteady and inertial dynamics of a small active particle in a fluid}
	\author{T. Redaelli}
	\affiliation{Aix Marseille Univ., CNRS, Centrale Marseille, IRPHE, F--13384 Marseille, France}
	\author{F. Candelier}
	\affiliation{ Aix-Marseille Univ., CNRS, IUSTI, F--13453 Marseille, France}
	\author{R. Mehaddi}
	\affiliation{ Universit\'e de Lorraine, CNRS, LEMTA, F--54000 Nancy, France}
	\author{B. Mehlig}
	\affiliation{Department of Physics, Gothenburg Universiity, SE--41296 Gothenburg, Sweden}

	\begin{abstract}
		It is well known that the reversibility of Stokes flow
		makes it difficult for small microorganisms to swim. Inertial effects break this reversibility, allowing new mechanisms of propulsion and feeding. Therefore it is important to understand the effects of unsteady and fluid inertia on the dynamics of microorganisms in flow. In this work, we show how to translate known inertial effects for non-motile organisms 
		to motile ones, from passive to active particles. The method relies on a principle used earlier by Legendre and Magnaudet (1997)
		to deduce inertial corrections to the lift force on a bubble from the inertial drag on a solid sphere, using the fact that small inertial effects are determined by
		the  far field of the disturbance flow.
		The method allows for example to compute the inertial effect 
		of unsteady fluid accelerations on motile organisms, and the inertial forces such organisms experience in steady shear flow. We explain  why the method fails to describe the effect of convective fluid inertia.
	\end{abstract}
	
	\maketitle
	
	\section{Introduction}
	\label{Intro}

	Active-matter research asks, amongst other questions, how small active particles such as motile micro-organisms move in a fluid, how they interact via hydrodynamic interactions, and how their motion is affected by obstacles, such as the walls that contain the fluid. It is usually assumed that the particles are so small that their dynamics is overdamped. In this steady creeping-flow or Stokes limit, fluid accelerations caused by the moving particle can be neglected,  and it is well known that the reversibility of Stokes flow makes it difficult for small microorganisms to swim \cite{purcell}.
	How does the particle dynamics change when the particles become larger, so that inertial effects begin to matter? When fluid inertia is weak,  one can compute the inertial dynamics by perturbation theory 
	around the creeping-flow limit  \cite{proudman1957,Sano81,Lovalenti93,childress1964,Gotoh90,saffman1956,Harper68,McLaughlin91,mei_adrian_1992,Asmolov99,Miyazaki95a,Candelier2019}.  Singular perturbation theory and asymptotic-matching methods have been used to investigate the effect of fluid inertia upon the motion of small passive particles. For example, convective fluid inertia enhances the drag in homogeneous \cite{oseen1910uber} and stratified \cite{Mehaddi2018} flows, and it causes a torque on non-spherical particles moving  in a spatially uniform flow \cite{Cox65,Kha89,Dab15,Can16}. Fluid accelerations due to shear result in lift forces \cite{saffman1965,Stone00,Candelier2019}. They modify the angular velocity of small particles in shear flow \cite{stone2016,Meibohm16}, and in turbulence \cite{Candelier2016}. 
	Even weak inertial forces can  have a substantial effect on the dynamics, for example if the unperturbed motion is marginally stable \cite{saffman1956,subramanian2005,einarsson2015a,herron1975sedimentation}.  
	The  results mentioned above apply in the steady limit. Unsteady fluid inertia  gives rise to
	the BBO history force \cite{boussinesq1885resistance,basset1888treatise,oseen1927neuere} 
	and added-mass effects~\cite{taylor1928forces,auton1988force,magnaudet1995accelerated}.

For active particles less is known. The rich biodiversity of phytoplankton and zooplankton  results in a substantial diversity of propulsion strategies. The associated inertial parameters span a wide range of values. Fluid inertia matters more for larger organisms, and for those with vigorous swimming gaits. Inertial effects break the time-reversibility of Stokes flow, and may thus provide new mechanisms of propulsion.
	\citet{Hamel7290} for example showed that the ciliate  \textit{paramecium}  makes use of inertial effects  to escape predators. 
	The clam \textit{clione antartica} exploits unsteady fluid inertia for propulsion \cite{Childress2004}. 
	Most of our present understanding of such fluid-inertia effects relies on ab-initio computer simulations, and upon perturbation theory in the relevant inertia parameter. 
	\citet{Lauga2007} calculated how much inertia is needed to allow for significant motion with reciprocal strokes. 
	\citet{Khair2014} and \citet{Wang2012a} computed the effect of convective inertia on a small spherical swimmer by perturbation theory in the particle Reynolds number. 
	 The ab-initio simulations of  \citet{Chisholm2016} confirm these results, and demonstrate how the theory fails for larger particle Reynolds numbers. 
	\citet{Ishimoto2013, IshimotoPhDThesis} studied how unsteady inertial effects  on the dynamics of a microswimmer depend upon its swimming stroke.
	Convective fluid inertia also affects the amount of fluid
an active particle moves around \cite{Chisholm2019}.
	Finally, feeding strategies of many marine organisms rely on unsteady inertial effects \cite{Magar2005,Michelin2013}. 
	Last but not least, artificial swimmers are usually larger than millimetres, because it is challenging to construct reliable and controllable propulsion mechanisms for very small robotic swimmers. As a consequence, fluid inertia tends to matter for such self-propelled robots \cite{Djellouli2017,Dombrowski2018}.

The motion of {\em copepods}, in particular,  illustrates the importance of fluid inertia for the swimming of marine microorganisms. During their life cycle, these organisms grow and change their gait. The steady creeping-flow approximation may hold for early life stages \cite{qiu}, but not necessarily for grown organisms   \cite{WadhwaPhDThesis,Visser2011,Catton2007}, and certainly not for the large accelerations generated when these 
	organisms escape predators by jumping \cite{Visser2011,Jiang2011}, as mentioned above.
	For such jumps,  the unsteady term $\partial \ve w/\partial t$
	in the Navier-Stokes equations matters ($\ve w$  is the velocity of the disturbance flow created by the particle).  In the simplest case, the inertial effect of this unsteady acceleration
	is described by the unsteady time-dependent Stokes equation, resulting in the above-mentioned history force. 
	For time-dependent but spatially homogeneous flows, it is understood how weak convective fluid inertia changes the history force for passive particles  \cite{Sano81,Lovalenti93,mei_adrian_1992}. But it is not known in general how the different inertial effects compete to determine the unsteady motion of microswimmers.
	
	Wang \& Ardekani \cite{Wang2012b} solved the  time-dependent Stokes equation to describe how history effects
	delay the decay of the velocity of a copepod after a jump. 
	The authors  found that the history force is
	essentially that of a passive particle, with a kernel that decays as $t^{-1/2}$ (a consequence of the diffusion of vorticity). 
	Fitting the parameters of their theory, the authors found agreement with earlier measurements by Jiang {\em et al.} \cite{Jiang2011}, despite the fact that the Reynolds number for jumping copepods is not small. The agreement is surprising because one might expect convective inertia to cause  the kernel to decay more rapidly, as it does for passive particles \cite{Sano81,Lovalenti93,mei_adrian_1992}.  
	
	The  results of Wang \& Ardekani \cite{Wang2012b}  raise further questions. First, is there a principle dictating that the history force  for an active particle is   essentially that of a passive one? Second, how does the history force depend on the shape of the swimming particle? After all, not all motile microorganisms are spherical. Third, since natural environments are rarely perfectly quiescent, how do fluid-velocity gradients affect the inertial dynamics?
	Does a small swimmer
	in a shear flow experience a lift force, analogous to Saffman's lift force on a passive particle in shear?  Can we predict the motion of an active particle in 
	a rotating or in a stratified fluid, mirroring what a passive particle experiences? 
	Does fluid inertia give
	rise to drag reduction for a swimmer in an elongational flow? More generally, how important are unsteady effects for swimmers  in time-dependent, spatially inhomogeneous flow? 
	
	The point of the present article is to show that certain results regarding the inertial dynamics of active particles can be directly inferred from the corresponding results for passive particles, making use of a fundamental principle employed by Legendre \& Magnaudet to determine the inertial lift on a small bubble in shear flow  \cite{legendre1997note}.
	 This allows us to answer at least some
	of the above questions. 
	The remainder of this article is organised as follows. In Section \ref{sec:model} we briefly summarise  the standard squirmer model, an  idealised model for a motile microorganism. Section \ref{sec:as} recapitulates how  small inertial effects can be treated in perturbation theory. In Section \ref{sec:trick} we describe the fundamental principle that allows to translate known results on the inertial dynamics of passive particles to active ones. Section \ref{sec:example} describes an example, and answers some of the questions raised above. We also discuss that the principle fails under certain circumstances, and speculate about the possible reasons for this failure. 
	We summarise our results in the conclusions, Section \ref{sec:conclusions}.
	
	\section{Squirmers: models for swimming at low Reynolds numbers}
	\label{sec:model}
	Swimming at low Reynolds numbers is commonly studied using the squirmer model  originally developed by Lighthill \cite{Lighthill} and Blake \cite{Blake1971}. In the simplest version of this idealised model, the effect 
	of beating cilia  is represented in terms of a boundary condition, in the form tangential flow components on the surface of a spherical swimmer  \cite{Visser2011,Pedley2016a,Pak,Wang2012a,Chisholm2016}
		\begin{equation}
	\label{eq:vt}
	v_s(\theta) = B_1 \sin\theta + B_2 \sin\theta\cos\theta\,.
	\end{equation} 
	The dynamics is usually discussed
	in terms of the parameter 
	$\beta = B_2/B_1$, where $B_1$ is defined to be positive. One distinguishes pullers  ($\beta >0$) from pushers ($\beta < 0$), reflecting the form of the fluid disturbance caused by the swimmer. The first term in Eq.~(\ref{eq:vt}) has dipole symmetry, the second term is a stresslet  with quadrupolar symmetry. Higher-order contributions are neglected in Eq.~(\ref{eq:vt}), but may matter for artificial swimmers, for instance for 
	Janus particles \cite{Volpe2016,Bickel2013,Shen2018}.
	
	Blake's  solution for the disturbance flow  assumes a steady swimming velocity 
	$\dot{\ve x}$ (the dot stands for the time derivative) and makes use of the fact that 
	the total force on the particle vanishes  because drag must equal thrust for a steady swimmer \cite{Chisholm2016}.  
	In other words, to compute the steady swimming velocity $\dot{\ve x}$, one solves  the steady Stokes equation  for arbitrary $\dot{\ve x}$ to  find the hydrodynamic force \cite{Lighthill,Blake1971}
	\begin{equation}
	\label{eq:f0p}
	\ve f_{\rm h}^{(0)}  = 6 \pi \mu a \Big( \frac{2}{3}B_1 \ve n-\dot{\ve x} \Big)\,.
	\end{equation}
	Here the superscript  $^{(0)}$ denotes the leading-order steady Stokes contribution. Furthermore, $\ve n$ is the swimming direction of the spherical swimmer. Its radius is denoted by $a$, and $\mu$ is the dynamic viscosity of the fluid. The first term in Eq.~(\ref{eq:f0p}) is the active force produced by the beating cilia, 
	and $6\pi\mu a$ is the resistance coefficient of the spherical particle. For a passive particle, $B_1=0$, Eq.~(\ref{eq:f0p}) reduces to  force on a passive sphere. 
	Setting $\ve f_{\rm h}^{(0)}=\ve 0$  in Equation(\ref{eq:f0p}) yields the  swimming velocity
	\begin{align}
	\label{eq:V0}
	\dot{ \ve x} = \tfrac{2}{3} B_1 \ve n
	\end{align}
	in the steady Stokes limit. This expression is independent of $\beta$ or any higher multipole corrections to Eq.~(\ref{eq:vt}).
	
	Oseen corrections to Eq.~(\ref{eq:V0}) describe how steady convective fluid inertia  affects the steady swimming speed. 
		Khair {\em et al.} \cite{Khair2014} found that
		\begin{align}
		\label{eq:VRe}
		\dot{\ve x} = \tfrac{2}{3} B_1 \ve n \Big(1-\frac{3\beta}{20}\rm{Re}_p\Big)\,.
		\end{align}
		Here Re$_p$ is the particle Reynolds number, measuring the effect of steady convective fluid inertia (see Section \ref{sec:as} for the non-dimensional parameters of the problem).
		Equation (\ref{eq:VRe}) corrects an earlier result of Wang and Ardekani \cite{Wang2012a} who obtained the value $0.11$  for the numerical coefficient
		in front of  Re$_p$, instead of $3/20$.  Khair {\em et al.} \cite{Khair2014} also computed the ${\rm Re}_p^2$-correction.
		Their results are  in good agreement with those of numerical simulations \cite{Chisholm2016}.
	
	Strictly speaking, beating cilia cause  time-dependent perturbations, resulting in time-dependent coefficients $B_l(t)$ 
	in Eq.~(\ref{eq:vt}) and thus a time-dependent swimming speed $\dot{\ve x}(t)$. To take into account such time-dependent boundary conditions, Wang \& Ardekani \cite{Wang2012b} solved the time-dependent Stokes equation for a spherical squirmer, neglecting the effect of convective fluid inertia. They showed that the unsteady acceleration in the time-dependent Stokes equation gives rise to a history force, much like the
	history force \cite{boussinesq1885resistance,basset1888treatise,oseen1927neuere} experienced by a passive sphere  in a quiescent fluid.
	
	\section{Singular perturbations and their non-dimensional parameters}
	\label{sec:as}
	The swimmer accelerates the surrounding fluid as it moves,
	and this causes the fluid-inertia correction in Eq.~(\ref{eq:VRe}), as well as the history force mentioned above.
	When such inertial corrections are small, one can use perturbation theory to analyse their effects.
	To determine the relevant  non-dimensional parameters, one  expresses the Navier-Stokes equations for the disturbance velocity in non-dimensional form. It is customary to non-dimensionalise lengths by the size of the swimmer (its radius $a$, say), time by a characteristic time $\tau_{\rm c}$ over
	which the slip velocity varies over surface of the swimmer, velocities  by a characteristic slip  velocity $u_{\rm c}$, and
	fluid-velocity gradients by a characteristic strain rate $s_c$. 
	This leads to three independent non-dimensional parameters
	\begin{equation}
	{\rm Re}_p =  \frac{a u_c}{\nu}\:, \quad {\rm Re_s}= \frac{a^2 s_c}{\nu}\:,\quad {\rm Sl} = \frac{a}{u_c \tau_c}\,.
	\end{equation}
	Here $\nu=\mu/\rho$ is the kinematic viscosity of the fluid with mass density  $\rho$. First, the particle Reynolds number  ${\rm Re}_p$
	describes how convective terms based on the slip-velocity  affect the dynamics of the fluid disturbance.
	This is the Oseen problem. Convection can make a qualitative difference because it can carry the disturbance away from the swimmer  more rapidly than diffusion alone. 
	Second,   for swimmers in spatially inhomogeneous flows, the shear Reynolds number ${\rm Re}_s$ quantifies how the imposed  fluid-velocity gradients affect the disturbance caused by the swimmer (Saffman problem). Third, the magnitude of unsteady effects is determined by the product  ${\rm Re}_p  {\rm Sl}=a^2/(\nu \tau_c)$ where $\rm{Sl}$ is the Strouhal number.  Unsteady fluid accelerations are described by the time-dependent Stokes equation. Nevertheless, they are essentially an inertial effect, describing how the hydrodynamic forcing decays after a perturbation. 
	
	From now on we use the non-dimensional variables described above, and expand the hydrodynamic force
in a small 
	parameter $\varepsilon$:
	\begin{equation}
	\label{eq:expansion}
	 \ve f_{\rm h} = \ve f^{(0)}_{{\rm h}} + \varepsilon  \ve f^{(1)}_{\rm h}+\ldots .
	\end{equation}
	When convective effects dominate the inertial correction, the perturbation parameter is 
	$\varepsilon = {\rm Re}_p$. When fluid-velocity gradients  are more important, 
	the parameter is instead $\varepsilon = \sqrt{{\rm Re}_s}$. When   unsteady effects 
	matter most, $\varepsilon=\sqrt{{\rm Sl}\,{\rm Re}_p}$ is the natural parameter. 
	In general it is challenging to compute the first inertial correction, $\varepsilon  \ve f^{(1)}_{\rm h}$, in Eq.~(\ref{eq:expansion}).
	Regular perturbation theory fails when
	perturbations which are small close to the particle dominate far from it. In this case, singular perturbation theory is required. Masoud \& Stone \cite{Masoud2019} used
	the reciprocal theorem to compute such inertial corrections (a particular example is given in the appendix of Ref.~\cite{Khair2014}). This approach  relies upon the fact that the disturbance flow decays as $r^{-2}$ far from the swimmer, faster than the Stokeslet which decays as $r^{-1}$. This  follows from the fact that the lowest-order dynamics is obtained by setting $\ve  f_{\rm h}^{(0)}$ to zero, and it
	allows to 
	apply the reciprocal theorem in its simplest form.
	
	Here we use the method of matched asymptotic expansions instead, a standard scheme to compute small fluid-inertia effects on the dynamics of passive particles in a fluid. For a passive sphere in uniform flow, for example, Proudman \& Pearson \cite{proudman1957} computed  the correction  $\ve f_{\rm h}^{(1)}$ due to convective fluid inertia, obtaining the Oseen correction \cite{vesey2007simple} to Stokes force,
	with $\varepsilon = {\rm Re}_p$.  Childress \cite{childress1964}   computed the hydrodynamic force on a light sphere rising in a solid-body rotating fluid, to order $\varepsilon$. Saffman \cite{saffman1956} derived the lift force on a passive sphere in shear flow to order $\varepsilon= \sqrt{{\rm Re}_S}$. Note that the time-dependent term $\partial\ve w/\partial t$ in the time-dependent Stokes equation is an inertial 
	perturbation to the steady Stokes equation. This term gives rise to the BBO history force at order $\varepsilon=\sqrt{{\rm Re}_p{\rm Sl}}$. So the history force is essentially an inertial effect, just like the inertial Oseen and Saffman corrections to the hydrodynamic force.
	The advantage of matched asymptotic expansions is that it allows to translate much of what is known for passive particles to active swimmers, as we show below.  
	
	\section{A fundamental principle of asymptotic matching}
	\label{sec:trick}
	The method of matched asymptotic expansions can be used when the disturbance flow $\ve w$ produced by the swimmer is given
	by Stokes equation plus  inertial corrections  that are small near the particle.
	In this inner region, the disturbance velocity can simply be  expanded in the small parameter $\varepsilon$. 
	Far from the particle, in the outer region, the perturbation may be substantial. In this case 
	regular perturbation theory fails.
	To obtain the outer solution, far from the particle, the boundary conditions on the particle surface are approximated by a source term  in the form of Dirac $\delta$-function \cite{Schwartz}. As a consequence, the outer problem can be solved in closed form by Fourier transform. Inner and  outer solutions are  matched in the matching region, at $r \sim \varepsilon^{-1}$ \cite{saffman1956}. 
	In practice this means that
	one needs to solve the inner problem with an $\varepsilon$-dependent boundary condition in the matching region \cite{Meibohm16}, determined by the $\varepsilon$-dependent outer solution. 
	Note that this solution may
	depend on time if the boundary conditions on the surface of the particle are time dependent, such
	as the time-dependent coefficients $B_l(t)$ mentioned in Section \ref{sec:model}.  
	
	The calculations are greatly simplified because the boundary condition for the inner solution far from the particle is just a uniform flow
	\cite{saffman1956}. In order to illustrate how this  uniform flow may arise, we disregard the details and consider only the schematic form of an outer solution, namely
	\begin{align}
	\label{eq:wrt1}
	w(r,t) = \frac{\mbox{exp}{[-\varepsilon \mathcal{U}(t) r]}}{r}\,.
	\end{align}
	Here $r$ is the distance from the particle,
	and $\mathcal{U}(t)$  is a time-dependent velocity. Recall that Eq.~(\ref{eq:wrt1}) is written in non-dimensional form (Section \ref{sec:as}). 
	For small values of $\varepsilon$, Eq.~(\ref{eq:wrt1}) can be expanded as 
	\begin{align}
	\label{eq:wrt_expansion}
	w(r,t) = \frac{1}{r} - \varepsilon \: \mathcal{U} + \mathcal{O}\left(\varepsilon^{2} r \right) + \ldots\,.
	\end{align}
	In this expansion, the first term $r^{-1}$ corresponds to the Stokeslet, 
	decaying as $r^{-1}$ far from the particle.  The presence of this slowly decreasing  term makes the perturbation problem singular. 
	The next term, of order $\varepsilon$,  is the uniform flow mentioned above, a uniform velocity that does not depend on the spatial coordinate $r$. This term becomes an
	outer boundary condition for the inner problem that must be solved to obtain the  $\varepsilon$-correction to the hydrodynamic force.
	Note that all terms in Eq.~(\ref{eq:wrt_expansion}) are of the same order in the matching region, for $r \sim \varepsilon^{-1}$, which reveals that the problem is singular.

	An important feature of the uniform boundary condition at order $\varepsilon$ is that it depends on the shape of the swimmer in a simple way. 
	The shape
	of the swimmer is  encoded in the amplitude of this $\delta$-function mentioned above, in terms of the Stokes resistance tensor $\ma R$  of the swimmer \cite{Candelier2019}, and through  the known  propulsive force produced by the particle in Stokes limit, the first term in Eq.~(\ref{eq:f0p}). As a consequence,  only the resistance tensor  $\ma R$ and the propulsion force are needed to determine the leading-order inertial correction to the hydrodynamic force. This  elegant trick  appears to be due to  Legendre \& Magnaudet \cite{legendre1997note} who used it to determine the lift force on a bubble in  shear  flow. In short, the resistance tensor of a solid sphere is  $\ma R= 6\pi \ma I$, 
	with identity tensor $\ma I$, whereas  the resistance tensor  for a bubble is smaller by a factor of  $\tfrac{2}{3}$.
	This implies that the amplitude of the uniform flow far from the bubble equals  $\tfrac{2}{3}$ that of a solid sphere. 
	This uniform flow in turn generates an inertial correction of the same form as the Saffman lift force on a solid sphere, 
	but scaled by a factor of  $\tfrac{2}{3}$. Taken together this means that  the inertial lift force on a bubble equals that for a solid sphere multiplied by a factor $\tfrac{4}{9}$, to order $\varepsilon$. Harper \& Chang \cite{Harper68} used related considerations to determine the lift force on a dumbbell
	in a shear flow. 
	More recently, the same principle was employed by Candelier, Mehlig \& Magnaudet \cite{Candelier2019} to determine inertia effects on a passive particle of arbitrary shape in steady linear flow. 
	
	\section{Inertial corrections to the hydrodynamic force on an active particle}
	\label{sec:example}
	The main point of the present paper is to note that the principle outlined in the previous Section can be applied to active particles. This  allows to 
	infer inertial corrections to the hydrodynamic force and torque on an active particle moving through a fluid. 
	As we will show, the principle applies provided that the equations for the disturbance flow close to the swimmer
	are essentially Stokes equations, perturbed by small terms that scale as  $\varepsilon^{n}$ with $n>1$. 
	More precisely, consider an inertial problem of the form
	\begin{equation}
	\label{eq:linear_operator}
	-\ve \nabla p + \Delta\ve w= {\mathscr{L}(\ve w , \varepsilon)}\,,
	\end{equation} 
	a Stokes problem for the disturbance velocity $\ve w$, perturbed by an inertial correction
	$\mathscr{L}(\ve w,\varepsilon)$
	with perturbation parameter $\varepsilon$. An example is the Saffman problem, describing a particle moving with a steady velocity in a time-independent linear undisturbed flow with fluid-velocity gradients $\ma A$. In this case,
	$\mathscr{L}(\ve w , \varepsilon)=\varepsilon^2(\ma{A}\cdot\ve w + (\ma{A}\cdot\ve r)\cdot\nabla\ve w)$, 
	and the perturbation parameter is given by $\varepsilon = \sqrt{{\rm Re}_s}$. A second example is the unsteady Stokes equation,
	where $\mathscr{L}(\ve w , \varepsilon)=\varepsilon^2\partial \ve w/\partial t$ with $\varepsilon = \sqrt{{\rm Sl}{\rm Re}_p}$.
	Now, if the perturbation is of the order
	\begin{equation}
	\label{eq:condition}
	\mathscr{L}\sim\varepsilon^{n} \quad \mbox{with}\quad n>1\,,
	\end{equation}
	then the two first terms of the inner expansion -- the first two terms in the expansion of Eq.~(\ref{eq:linear_operator}) in the parameter $\varepsilon$ -- are both solutions of the steady Stokes equation.
	This is the key fact that allows us to apply the principle. 
	
	As an example, consider the second case,  the unsteady inertial perturbation.
	Since the perturbation parameter is $\varepsilon^2$, one might expect that only terms even in $\varepsilon$ occur in the perturbation expansion for the disturbance velocity. For singular perturbation problems, however, other terms may appear: in the present case terms that are odd
		in $\varepsilon$, in other cases even terms that contain $\log\varepsilon$ \cite{Hinch1995}. This is well known, and such additional terms are sometimes
		called \lq switchback\rq{} terms. Our point here is that the leading switchback term, 
		of order $\varepsilon$, has a very simple form.
	
	In order to flesh out these arguments, consider first the expansion of the disturbance velocity close to the particle
		\begin{equation}
		\label{inner_exp}
		\ve w_{\rm in} = \ve w_{\rm in}^{(0)} + \varepsilon \ve w_{\rm in}^{(1)} +\ldots
		\end{equation}
		Here $\ve w_{\rm in}^{(0)}$ obeys Stokes equation with boundary condition  $\ve w_{\rm in}^{(0)} = \dot{\ve x}+ v_{\rm s}\ve n$ on the surface of the swimmer.
		Undercondition (\ref{eq:condition}), the first-order correction $\ve w_{\rm in}^{(1)}$ too obeys Stokes equation, but now with boundary condition $\ve w_{\rm in}^{(1)} =\ve 0$ on
		the swimmer surface. To solve these two Stokes problems, boundary conditions far from the swimmer are needed. They are obtained by
		matching the inner solution to an outer solution for the disturbance velocity, evaluated in the matching region. This outer solution, $\ve w_{\rm out}(\ve r,t)$, 
		is obtained by solving the full inertial problem (\ref{eq:linear_operator}), including the perturbation $\mathscr{L}$. This becomes possible if one replaces the boundary condition on the surface of the swimmer by a source term
		that represents the presence of the swimmer. This source term is of the form ${\ve D}(t) \:\delta$, where $\delta$ is the Dirac delta function, and the amplitude $\ve D$ remains to be determined. To this end, $\ve w_{\rm out}$ is  expanded as 
		\begin{equation}
		\ve w_{\rm out}({\ve r},t)= \ve{\mathcal{T}}^{(0)}({\ve r},t) + \varepsilon \ve {\mathcal{T}}^{(1)}(t) + \ldots
		\label{outer_exp}
		\end{equation}
		To lowest order, the outer solution is  \cite{Meibohm16}
		\begin{equation}
		\label{order_0}
		\ve{\mathcal{T}}^{(0)} ({\ve r},t) =\ma{G} (\ve r)
		\ve D(t) \quad \mbox{with} \quad [\ma{G}]_{i j} = \frac{1}{8\pi} \left(\frac{\delta_{ij}}{r} + \frac{r_i r_j}{r^3} \right)\,.
		\end{equation}
		When condition  (\ref{eq:condition}) holds, the first order is given by \cite{Meibohm16}
		\begin{equation}
		\label{eq:U}
		\ve{\mathcal{T}}^{(1)}(t) = - \int_0^t \!\!\mbox{d}\tau\,\ma{K}(t-\tau) \frac{\mbox{d} \ve D{(\tau)}}{\mbox{d}\tau }\equiv \ve{\mathcal{U}}(t)  \,,
		\end{equation}
		where it is assumed that
$\ve D(0) =\ve 0$ \cite{Candelier2019}, otherwise the lower bound of the integral extends to $-\infty$.
		It is important to note that  $\ve{\mathcal{T}}^{(1)}(t)$ is a uniform flow,  $\ve{\mathcal{U}}(t)$, just like the second term in Eq.~(\ref{eq:wrt_expansion}).
	The nature of the inertial perturbation, $\mathscr{L}$, is encoded in the temporal behaviour of the kernel  $\ma{K}(t)$ which does not depend on particle shape. The kernel is known for the Saffman problem (see Ref.~\cite{Candelier2019} and references cited therein), when the undisturbed problem
		is a shear, solid-body rotation, or a two-dimensional elongational flow. For the unsteady problem,  $\ma K(t)$ is just the BBO kernel $\ma{I}/[6 \pi \sqrt{\pi t}]$ \cite{boussinesq1885resistance,basset1888treatise,oseen1927neuere}.
	
	Equations (\ref{order_0}) and (\ref{eq:U}) provide the desired boundary conditions for $\ve w_{\rm in}^{(j)}$, namely
		$\ve w_{\rm in}^{(j)}=\ve{\mathcal{T}}^{(j)}$ at $r\sim \varepsilon^{-1}$ for $j=0$ and $1$. Consider first the Stokes problem at order $\varepsilon^0$. Since  $\ve w_{\rm in}^{(0)}$
		is the Stokes disturbance velocity, we infer that $\ve D$ must equal  the Stokes force exerted by the swimmer upon the fluid:
	\begin{equation}
	\label{eq:f0}
	\ve D(t) = -\ve{f}_{\rm h}^{(0)}(t) =-[ \ve f_{\rm{a}}(t)- {\ma{R}}(t)  \dot{\ve x }(t)]\,.
	\end{equation}
	Eq. (\ref{eq:f0}) reflects the fact that 
	the velocity  over the surface of the organism can be written as a sum of 
	two terms. The first one  is the active
	contribution \cite{Lauga2009}, written in terms of  the active force produced by the swimmer,  $\ve f_{\rm{a}}{(t)}$.
	The second term  
	is the passive part of the disturbance force. It depends upon the instantaneous shape of the particle encoded in the resistance tensor ${\ma{R}}(t)$. 
	
	Since $\ve{\mathcal{U}}(t)$ is a uniform flow, the order-$\varepsilon$ Stokes problem for $\ve w^{(1)}_{\rm in}$ is just the Stokes problem for a frozen particle (a particle kept at fixed position and orientation in the uniform  flow $\ve{\mathcal{U}}{(t)}$ at infinity). The desired  inertial correction $\varepsilon \ve f^{(1)}_{\rm h}{(t)}$ is thus  simply given by the hydrodynamic force such a particle experiences,
		\begin{equation}
		\label{eq:f1}
		\ve f_{\rm h}^{(1)} (t)= 
		{\ma{R}}(t)   \ve{\mathcal{U}}(t)\,.
		\end{equation}
		Taking Equations  (\ref{eq:U}),  (\ref{eq:f0}), and  (\ref{eq:f1}) together we find the following expression for the inertial correction to the hydrodynamic force for an active particle:
	\begin{equation}
	\label{eq:trick_result}
	\ve f_{\rm h}^{(1)}{(t)}  =    {\ma{R}}(t) \int_0^t \!\!\mbox{d}\tau\,\ma{K}(t-\tau) \frac{\mbox{d}}{\mbox{d}\tau } \left[\ve{f}_{\rm{a}}(\tau) - {\ma{R}}(\tau) \dot{\ve x}(\tau) \right] \,.
	\end{equation}
	This is the main result of the present paper. Note that the active part  $\ve f_a{(t)}$ in  $\ve f^{(0)}_{\rm h}{(t)}$ contributes 
	linearly to the inertial correction at order $\varepsilon$, whereas the resistive part appears in the form
	of two possibly time-dependent  factors ${\ma R}(t)$,  one inside the integral, and one  outside. 
	
	 As an example, consider how to obtain the history force upon a spherical squirmer
	moving with velocity $\dot{\ve x}(t)$ in  direction $\ve{n}(t)$ in a fluid at rest \cite{Wang2012b}. Assuming that convective and shear inertia are negligible, 
	Stokes equations are perturbed only by the unsteady term. In this case, the small parameter is  $\varepsilon = \sqrt{a^2 /(\nu \tau_c)}$ and the uniform flow is 
	given by
	\begin{equation}
	\label{eq:U2}
	\ve{\mathcal{U}}{(t)}= \frac{1}{6\pi}\int_0^t \!\!\mbox{d}\tau\, \frac{1}{\sqrt{\pi (t-\tau)}} \frac{\mbox{d} \ve f_{\rm h}^{(0)}{(\tau)}}{\mbox{d}\tau} \,.
	\end{equation}
	The  $t^{-1/2}$-decay of the kernel describes how disturbance-velocity gradients relax due to viscous diffusion \cite{boussinesq1885resistance,basset1888treatise,oseen1927neuere}.
	The resistance tensor is simply that of a sphere, $6\pi \ma{I}$, and the active force is $\ve f_{\rm{a}}{(t)} =6\pi\frac{2}{3}B_1(t) \,\ve n(t)$.  Using Eqs.  (\ref{eq:expansion}), (\ref{eq:trick_result}), and (\ref{eq:U2}), immediately gives 
	\begin{equation}
	\label{eq:U3}
	\ve f_{\rm h} {(t)}= - 6 \pi [\dot{\ve{x}}{(t)} - \frac{2}{3}B_1{(t)} \ve{n}{(t)}]- 6\pi\varepsilon   \int_0^t\!\!\mbox{d}\tau \, \frac{1}{\sqrt{\pi (t-\tau)}} \frac{\mbox{d}}{\mbox{d}\tau}\big[\dot{\ve{x}} (\tau) - \frac{2}{3}B_1(\tau) \ve{n}(\tau)\big]
	\end{equation}
	for the unsteady  hydrodynamic force on a squirmer (the added-mass  term is a higher-order contribution to the force of order $\varepsilon^2$). In the special case where $\ve n$ is a constant vector,  Eq. (\ref{eq:U3}) reduces to a result obtained earlier by  Wang \& Ardekani \cite{Wang2012b} using the method of Laplace transforms. 
		This expression was first derived by Morrison \cite{morrison}, describing the electrophoretic force upon a dielectric sphere
		in a transient electric field.
	
	Eq.~(\ref{eq:trick_result}) explains why the hydrodynamic force on an unsteady spherical squirmer must be of the form (\ref{eq:U3}), and it demonstrates how to generalise the result
	to  arbitrary shapes.  
	Since motile aquatic microorganisms exhibit  a rich variety of shapes and swimming gaits \cite{childress1981,beckett1986}, 
	it is useful that Eq.~(\ref{eq:trick_result}) separates the contributions to the hydrodynamic force due to particle shape, through the tensor {$\ma{R}$}, from those due to
	the time-dependence of the  fluid disturbance, through the kernel $\ma{K}$.  
	
	Eq.~(\ref{eq:trick_result}) implies that the inertial correction vanishes in the steady limit, for example for a steady squirmer with $\dot{\ve x} = \tfrac{2}{3} B_1 \ve n$.
		In this case $\ve f^{(0)}_{\rm h}$ vanishes, so that there is no inertia-induced force at order $\varepsilon$ (see for example Refs.~\cite{choudhary2021how,shaik2021squirming}).
		As a consequence, the disturbance flow produced by the swimmer is a stresslet. For transient dynamics, on the other hand, 
		$\ve f^{(0)}_{\rm h}{(t)}$  is non-zero, resulting in  transient inertial corrections. Such corrections arise for sudden starts or stops, and also when $B_1{(t)}$  is a smooth
		but rapidly varying time-dependent function, so that a difference 
		between $\dot{\ve x}(t)$ and $\tfrac{2}{3} B_1(t){\ve n(t)}$ is maintained.
	
	The above example concerns unsteady inertia effects, and answers  two questions raised in the Introduction, namely why is the history force is the same for active and passive particles?
	How does it depend on the shape of the swimmer?
	There are several other situations where Eq.~(\ref{eq:trick_result}) may help to determine the influence of fluid  inertia on the dynamics of active particles, for example	
	 active particles in shear flows. 
	The amoeba \textit{Naegleria Fowleri}, for example, thrives in water-discharge flows
	from industrial plants which can have shear rates as high as $85 000\,$s$^{-1}$ \cite{perrin2018design}. For a micron-sized organism
	this results in a shear-Reynolds number of the order of ${\rm Re}_s\sim 0.1$, so that $\varepsilon \sim 0.3$, which means that  fluid inertia should matter.  The equation of motion  for a spherical squirmer in a shear flow follows directly from Eq. (\ref{eq:trick_result}). In non-dimensional form it reads:
		\begin{equation}
		R \frac{4\pi}{3} \varepsilon^2 \ddot{\ve x} {(t)}
		= -6 \pi \Big[\dot{ \ve x}(t) - {\ve U^\infty} ({\ve x}) - \frac{2}{3} B_1(t) \ve {n}(t) \Big]
		- 6 \pi \varepsilon \int_0^t \ma{K}(t-\tau) \frac{\mbox{d}}{\mbox{d} \tau} \Big[ \dot{ \ve x}(\tau) - {\ve U^\infty} ({\ve x}) - \frac{2}{3} B_1(\tau) \ve {n}(\tau)\Big] \mbox{d}\tau + \ve F_{\rm{ ext}}\,.
		\label{eq:motion_Shear}
		\end{equation}
	Here
		$R$ is the particle-to-fluid density ratio, ${\ve U^\infty} ({\ve x})$ is the undisturbed shear flow  at distance $\ve x$ from the particle centre \cite{Candelier2019}, and $\ve F_{\rm{ ext}}$ is an external force. 
		 In the steady limit, when  $B_1 \ve n$ tends to a steady vector,  the velocity of the particle approaches
		\begin{equation}
		\label{eq:shearxdot}
		\dot{\ve x} = \ve U^\infty({\ve x}) + \frac{2}{3} B_1 {\ve n} + \frac{1}{6\pi} \ve F_{\rm{ ext}} - \varepsilon   \frac{1}{6\pi} \bar{\ma{K}} \cdot \ve F_{\rm{ ext}} \,.
		\end{equation}
		Here $\bar{\ma{K}} $ is the steady-state limit of the kernel. In this limit, the $\varepsilon$-inertia correction induced is the same as for a passive particle. 
		However, the time required to reach the steady state is usually very long (typically much larger than the inverse shear rate). During this transient, the inertial effects on passive and active particles differ.
		The last term of Eq.~(\ref{eq:shearxdot}) shows that an active particle moving with a steady velocity in a shear flow may experience a lift force, but only if there is an external force. Also note that if $B_1(t) \ve n(t)$ varies rapidly, then $\ma{K}$ simplifies to the BBO-kernel. In this limit, convective inertia  effects induced by shear do not matter.

	Finally, Eq.~(\ref{eq:trick_result}) can be used to describe active particles in density-stratified fluids, provided that convective inertia does not matter.  Eq.~(\ref{eq:trick_result}) shows that there may be important inertial corrections even for ${\rm Re}_p=0$, when the particle density exceeds the fluid reference density, resulting in an external force $\ve F_{\rm ext}$ due to gravity.
	In Refs.~\cite{more_ardekani_2020,doostmohammadi2012low}, by contrast, inertial effects due to convective fluid inertia were considered. As explained below, the principle does not apply in this case.

	Last but not least let us discuss when and how the principle may fail.  We encountered one example where the principle does not work,  an active particle moving with a large enough slip velocity so that convective inertia cannot be neglected. This is the Oseen problem mentioned in Section \ref{sec:example}. In this case, the perturbation in the perturbed Stokes problem (\ref{eq:linear_operator}) takes the form 
	$\mathscr{L} = - \varepsilon \:\dot{\ve{x}} \cdot \boldsymbol{\nabla} \ve{w}$ in a quiescent fluid, with $\varepsilon = \rm{Re}_p$. 
	This perturbation is of order $\varepsilon$, so the above condition needed to justify the principle is not fulfilled. As a consequence,   the inner solution is no longer governed by a Stokes equation, as assumed above,  but by an inhomogeneous differential equation including terms arising from the solution at leading order in $\varepsilon$. 
	This explains why the inertial $\rm{Re}_p$-corrections to the force acting on a spherical squirmer described in Refs.~\cite{Wang2012a,Chisholm2016}  are not entirely recovered by the principle. Returning to the results of  Legendre \& Magnaudet \cite{legendre1997note}, note that the principle does give the correct Oseen correction to the drag force on a bubble: setting $\ve f_{\rm a}= {\ve 0}$ in Eq.~(\ref{eq:trick_result}), and inserting the resistance tensor of a spherical bubble,
	${\ma R} = 4\pi \ma I$, gives the correct result, namely that convective-inertial correction to the drag force on a bubble is $\tfrac{4}{9}$ times that for a solid sphere. We speculate that the result is correct for a spherical bubble because the  additional terms mentioned above vanish due to spherical symmetry.

	\section{Conclusions}\label{sec:conclusions}
	We described a fundamental principle that allows to infer the effect of fluid inertia on the dynamics of motile microorganisms moving in a fluid.
	In its original form, the principle was used by Legendre \& Magnaudet \cite{legendre1997note} to infer the lift force on a bubble in shear flow from Saffman's result for a solid particle.  Here we showed how the principle works for active particles. We demonstrated how it allows to infer the leading-order inertial corrections to the hydrodynamic force experienced by the particle, as it swims through the fluid. Our results explain why the history force 
	has essentially the same form for active and passive particles. We show that particle shape affects the 
	inertial correction to the hydrodynamic force in a very simple way, making it possible to investigate the effect of particle shape on the hydrodynamic force in a straightforward manner.  As an example we considered the effect of unsteady inertia on the hydrodynamic force. We also discussed other situations where the principle applies, namely shear flows and density-stratified fluids.
	
	Naturally, the method has limitations. As presented here, certain conditions must be fulfilled for the principle to work (Section \ref{sec:example}). These conditions are met for  unsteady and Saffman problems,
	but not for  the Oseen problem (convective fluid inertia). Therefore it remains an open problem how to compute convective inertial corrections to the history force on an active swimmer, even for a time-dependent  homogeneous flow. One question is how the convective corrections to the history force differ from those obtained by Lovalenti \& Brady  \cite{Lovalenti93} for a passive particle. At any rate, such convective effects must be considered to describe the disturbance velocities generated by jumping copepods, where unsteady and convective fluid inertia are likely to be equally important.

	\acknowledgments We thank an anonymous referee for pointing out Ref.~\cite{morrison} to us.  BM was supported by grants  from the Knut and Alice Wallenberg Foundation, grant numbers 2014.0048 and 2019.0079, and in part by VR grants  2017-3865 and 2021-4452. TR received funding from the European Research Council (ERC) under the European Union's Horizon 2020 research and innovation programme (grant agreement No 834238).

\end{document}